\documentclass[reprint,amsmath,amssymb,pra,twocolumn,longbibliography]{revtex4-2}
\usepackage{graphicx}
\usepackage{color}
\usepackage{bm}
\usepackage{amsmath}
\usepackage{csquotes}
 \usepackage{longtable}

 \def\prb{Phys. Rev. B}
 
 \def\pra{Phys. Rev. A}

\begin{document}

\title{Approximating transmission and reflection spectra near isolated nondegenerate resonances}

\author{Hongyao Wu}
\author{Lijun Yuan}
\email{ljyuan@ctbu.edu.cn}
\affiliation{School of Mathematics and Statistics, Chongqing Technology and Business University, Chongqing, 
China \\ Chongqing Key Laboratory of Social Economic and Applied Statistics, Chongqing Technology and Business University, Chongqing, China}

\author{Ya Yan Lu}
\affiliation{Department of Mathematics, City University of Hong Kong, Kowloon, Hong Kong, China} 
\date{\today}

\begin{abstract}
A linear scattering problem for which incoming and outgoing waves are
restricted to a finite number of  radiation channels can be precisely
described by a frequency-dependent scattering matrix.  The entries of
the scattering matrix, as functions of the frequency, give rise to the
transmission and reflection spectra. To find the scattering matrix
rigorously, it is necessary to solve numerically the partial
differential equations governing the relevant waves.
  In this paper, we  consider resonant structures with an isolated nondegenerate resonant
  mode of complex frequency $\omega_\star$, and show that for real frequencies
  near $\omega_0 = \mbox{Re}(\omega_\star)$, the transmission and reflection spectra
  can be approximated using only the scattering matrix at $\omega_0$
  and information about the  resonant mode. We also present a revised 
  temporal coupled-mode theory that produces the same
  approximate formulas for the transmission and reflection spectra. 
  Numerical examples for diffraction of plane waves by periodic
  structures are presented to validate our theory.  
\end{abstract}

\maketitle

\section{Introduction}

A scattering problem is concerned with finding the outgoing waves when
a given incident wave impinges upon a structure. If both incoming and
outgoing waves are restricted to a finite number of radiation
channels, the complete solution of any linear scattering problem is
given by a finite scattering matrix that maps the amplitudes of the
incoming waves to those of the outgoing waves. Typically, the incoming
and outgoing waves are time-harmonic waves and the scattering matrix
depends on the frequency. Entries of the scattering matrix, as
functions of the frequency, can be used to find the transmission and
reflection spectra. As first observed by Wood~\cite{wood}, transmission
and reflection  spectra often exhibit rapid variations with sharp
peaks and/or dips. In numerous applications,  
a peak and a dip appear in a narrow frequency range
forming an asymmetric line shape --- a phenomenon called Fano
resonance~\cite{fano41,hessel,popov86,fan03}. 
For structures without absorption loss and with a proper symmetry, the peaks and dips
can actually reach $100\%$ and $0$,
respectively~\cite{popov86,gipp05,shipman12,bykov15,kras19}. 
It is widely accepted that Fano resonance is the consequence of 
interference between a direct (non-resonant) passway and a 
resonance-assisted indirect pathway~\cite{fan03}. 
In photonics, Fano resonance has found many applications including 
filtering, sensing and  switching~\cite{miro10,zhou14,limo17,bogd19}. 

To find the scattering matrix rigorously, it is necessary to solve the
governing partial differential equation (PDE), such as the Maxwell's
equations for electromagnetic waves. Accurate numerical solutions for
a large frequency range are expensive to obtain and do not provide
much physical insight. To improve the understanding on resonant
scattering phenomena, it is desirable to derive analytic models for
scattering matrices and transmission/reflection spectra. A good
analytic model should reveal the most important physical phenomena and predict 
the peaks and dips in transmission/reflection spectra. The temporal
coupled-mode theory (TCMT) is a simple system (for the amplitudes of the
resonant modes  and incoming and outgoing waves) constructed by
considering  energy conservation, reciprocity and time-reversal
symmetry~\cite{haus84,fan02,fan03,fan04,wang18,zhao19}. 
Although it is not derived from the governing PDE, TCMT produces a
simple model for the scattering matrix and predicts the peaks and dips
accurately. To use the TCMT for any specific application, it is necessary to find
the  resonant mode and estimate the scattering matrix $C$ for the
direct passway. While the resonant mode can be solved from the
governing PDE, the scattering matrix $C$ cannot be solved rigorously. A
different modeling approach, first suggested by Popov {\it et
  al.}~\cite{popov86}, is to approximate the entries of the scattering
matrix by simple rational functions based on their poles and zeros in
the complex plane~\cite{popov86,nevi95,fehre02,blan16}. It is well-known that
the complex frequency of a resonant mode is a pole of the scattering
matrix. Each entry of the scattering matrix has its
own zeros and they are complex in general. In case a dip in a 
transmission or reflection spectrum is actually 0, the corresponding
entry in the scattering matrix has a real zero. Both poles and zeros
can be found by solving the governing PDE.

In this paper, we  first consider scattering problems with two
radiation channels. For a resonant structure with a nondegenerate resonant
mode of complex frequency $\omega_\star$, we derive a simple
approximation for the frequency-dependent scattering matrix based on
the scattering matrix at $\omega_0 = 
\mbox{Re}(\omega_\star)$.  The corresponding approximations to the
transmission and reflection spectra are accurate for real frequencies
near $\omega_0$, and predict the peaks and dips in the spectra very 
well. Moreover, the derived approximate scattering matrix can be used
to determine the zeros of the transmission and reflection
coefficients, and to reveal the conditions under which the
zeros are real. To support and supplement our theory on approximating
scattering matrices, we develop a revised TCMT for general
scattering problems. The original TCMT
gives rise to a  symmetric scattering matrix that depends on the
scattering matrix $C$ of the direct passway~\cite{fan03}. For
scattering problems where  the original and 
reciprocal waves propagate in different  radiation channels, the
scattering matrix is  in general non-symmetric. 
Our revised TCMT produces a model scattering matrix which is
non-symmetric in general, is independent of $C$, and is
consistent with the approximation derived directly.

The rest of the paper is organized as follows. In
Sec.~\ref{sec:Smatrix}, we recall the definitions and
properties of scattering matrices and resonant modes for
two-dimensional (2D) structures with a single  periodic direction. In
Sec.~III, we derive approximate formulas for general $2\times 2$
scattering matrices and related transmission/reflection spectra.  In
Sec.~IV, we present  a revised  TCMT and derive a simple model for
the scattering matrix.
For validating our theory, numerical examples involving periodic
arrays of cylinders are presented in Sec.~\ref{sec:examples} 
The paper is concluded with a brief discussion in
Sec.~\ref{sec:conclusions}.

\section{Periodic structures}
\label{sec:Smatrix}

In this section, we introduce scattering matrices and resonant modes
using a two-dimensional (2D) periodic structure as an
example. Although the theories developed in the next two sections are 
applicable to more general cases, they will be validated by numerical
examples involving periodic structures.  
We consider a lossless periodic structure that is
invariant in $z$, periodic in $y$ with period $L$, and sandwiched
between two identical homogeneous media given for $x > D$ and $x < -D$,
respectively, where $\{x, y, z\}$ is a Cartesian coordinate system.  
The dielectric function $\varepsilon(x,y)$ of the structure and the
surrounding media is real and satisfies 
\begin{equation}
\varepsilon(x,y) = \varepsilon(x,y+L)
\end{equation}
for all $(x,y)$ and $\varepsilon(x, y) = \varepsilon_0 \ge 1 $ for $|x| > D$. In particular, the periodic structure may be a periodic array of
dielectric cylinders as shown in Fig.~\ref{fig_stru} of Sec.~V.

For the $E$ polarization, the $z$ component
of a time-harmonic electric field,  denoted by $u$, satisfies the
following 2D Helmholtz equation
\begin{equation}
  \label{helm}
 \frac{ \partial^2 u}{\partial x^2}  + \frac{ \partial^2 u}{\partial y^2} + \left( \frac{\omega}{c} \right)^2
  \varepsilon(x,y) u = 0,  
\end{equation}
where the time dependence is $\exp (- i \omega {\sf t})$, $\omega$ is the
angular frequency, $i$ is the imaginary unit, ${\sf t}$ is the time
variable, and $c$ is the speed of light  in vacuum. For a real
frequency $\omega$ and a real $\beta$ satisfying
\begin{equation}
\label{one_channel} | \beta | < \frac{ \omega }{ c}
\sqrt{\varepsilon_0} < \frac{2 \pi}{L} - |\beta|, 
\end{equation}
we illuminate the periodic structure by plane waves with wavevectors $(\pm \alpha,
\beta)$ from left and right, respectively, where
\begin{equation}
  \label{defalpha}
   \alpha = \sqrt{ (\omega/c)^2 \varepsilon_0 - \beta^2}
\end{equation}
is positive. The total field in the left homogeneous medium can be written as
\begin{eqnarray}
\nonumber 
  && u(x,y)  =  b_1^+ e^{ i [ \beta y + \alpha (x+D)]} +   b_1^- e^{
    i [ \beta y    - \alpha (x+D)] } \\
  \label{uleft}  && \qquad + \sum_{j\ne 0} b_{1j}
  e^{ i  \beta_j y  + \tau_j  (x+D) }, \qquad x < -D, 
\end{eqnarray}
where $b_1^+$ is the amplitude of the left incident wave,
$b_1^-$ is the amplitude of the outgoing wave in the left homogeneous
medium, 
\begin{equation}
  \label{delbg}
  \beta_j = \beta + 2\pi j/L, \quad 
  \tau_j = \sqrt{  \beta_j^2 - (\omega/c)^2 \varepsilon_0}
\end{equation}
for  $j\ne 0$, $\tau_j$ is positive,  and $b_{1j}$ is the amplitude of the evanescent plane
wave  ($j$th diffraction order) that decays exponentially as $x \to
-\infty$. Similarly, the total field in the right 
homogeneous medium is given by
\begin{eqnarray}
\nonumber
 &&  u(x,y) =  b_2^+ e^{ i [ \beta y  - \alpha (x-D) ] } +
    b_2^- e^{ i [ \beta
        y +  \alpha (x-D)]}   \\
  \label{uright}  && \quad + \sum_{j\ne 0} b_{2j}
  e^{ i   \beta_j y  -  \tau_j (x-D) }, \qquad x > D, 
\end{eqnarray}
where $b_2^+$ is the amplitude of the right incident
wave, $b_2^-$ is the amplitude of the right outgoing wave, $b_{2j}$
is the amplitude of the $j$th diffraction order that decays exponentially
as $x \to +\infty$.
Since the problem is linear, there is a $2\times 2$
matrix $S$, the scattering matrix, such that
\begin{equation}
  \label{Smatrix}  
  \left[  \begin{matrix}  b_1^{-} \\ b_2^{-} \end{matrix} \right]
  = S \left[  \begin{matrix}  b_1^{+} \\ b_2^{+} \end{matrix}
  \right], \quad
  S = \left[ \begin{matrix} r  & \tilde{t}\,  \\ t &
      \tilde{r}  \end{matrix} \right].   
\end{equation}
In the above, $r$ and $t$ ($\tilde{r}$ and $\tilde{t}$) are the reflection and  
transmission coefficients respectively, for left (right) incident
waves. 

It is clear that  $S$ depends on both $\omega$ and $\beta$. By
analytic continuation, the definition of $S$ can be extended to the
complex $\omega$ plane~\cite{popov86}.
Notice that for a complex $\omega$,
$\alpha$ and $\tau_j$ are also complex. 
Since we assume the structure is
lossless (i.e. $\varepsilon$ is real),  the power carried by
the incident and outgoing waves must be the same. This implies that for
real $\omega$ and $\beta$, $S(\omega, \beta)$ is a unitary
matrix~\cite{popov86}. The generalization to complex $\omega$ is 
\begin{equation}
\label{unitarity} 
S(\omega, \beta) S^* (\overline{\omega}, \beta) = I
\end{equation}
where $\overline{\omega}$ is the complex conjugate of $\omega$,
$ S^*(\overline{\omega}, \beta)$ is the conjugate transpose of $
  S$ evaluated at $(\overline{\omega}, \beta)$, and $I$ is the
identity matrix. A proof for Eq.~(\ref{unitarity}) is given
in Ref.~\cite{yuan19}.

Another important property of $S$ is
\begin{equation}
  \label{recip}
S^{\sf T}(\omega, \beta) =   S(\omega, -\beta), 
\end{equation}
where $S^{\sf T}$ is the transpose of $S$. This is a
consequence of the reciprocity and it is valid even when $\omega$ is
complex~\cite{popov86}. A proof can be found in Ref.~\cite{yuan19}.
Notice that, if $\beta \ne 0$,  the scattering matrix $S$ is non-symmetric in general.

For periodic structures with a proper symmetry, the scattering matrix
can be further simplified~\cite{popov86}. 
If the structure is symmetric in $y$, i.e. $\varepsilon(x,y) =
\varepsilon(x,-y)$, then $S$ is a symmetric and 
$t =\tilde{t}$. If the structure has an inversion symmetry,
i.e., $\varepsilon(x,y) = \varepsilon(-x,-y)$, then
$r=\tilde{r}$. Moreover, if the periodic structure is symmetric in $x$,
i.e., $\varepsilon(x,y) = \varepsilon(-x,y)$, then both reflection and
transmission coefficients for left and right incident waves are
identical, i.e., $t=\tilde{t}$ and $r=\tilde{r}$. More details can be found in
Refs.~\cite{popov86} and \cite{yuan19}. 

Different kinds of eigenmodes can exist in the periodic structure. Due
to the periodicity in $y$, any eigenmode is a Bloch mode given
by $u(x,y)
= e^{ i \beta y} \phi(x,y)$, where $\beta \in (-\pi/L, \pi/L]$ is the Bloch wavenumber and
$\phi$ is periodic in $y$ with period $L$. Moreover, an eigenmode must satisfy proper
boundary conditions as $x \to \pm \infty$. Typically, the wave field should decay
exponentially or be outgoing (radiating out power) as $x \to \pm
\infty$. In a lossless structure (without material loss), an eigenmode
that radiates out power to infinity ($x = \pm \infty$) cannot have
both real $\omega$ and real $\beta$. A resonant mode is an eigenmode
with a real $\beta$ and a complex $\omega$ satisfying the outgoing
radiation condition as $x \to \pm \infty$~\cite{fan02,amgad}. For the assumed time
dependence $e^{- i \omega {\sf t}}$, the imaginary part of $\omega$
is negative, and thus the amplitude of the resonant mode decays with
time  ${\sf t}$ .  If we assume
condition (\ref{one_channel}) is valid with $\omega$ replaced by
$\mbox{Re}(\omega)$, then a resonant mode
satisfies
\begin{equation}
  \label{rmode1}
  u(x,y) =
  d_1e^{i [ \beta y - \alpha (x+d)]} + \sum_{j\ne 0} d_{1j}\, 
  e^{i \beta_j y + \tau_j (x+d)}  
\end{equation}
for $x < -d$ and
\begin{equation}
  \label{rmode2}
  u(x,y) =
  d_2 e^{ i [ \beta y + \alpha  (x-d) ] } + \sum_{j\ne 0}
  d_{2j}\,   e^{ i \beta_j y - \tau_j (x-d)}
\end{equation}
for $x > d$, where $\alpha$ and $\tau_j$ are complex scalars satisfying
$\mbox{Re}(\alpha) > 0$, $\mbox{Im}(\alpha) < 0$,
$\mbox{Re}(\tau_j) > 0$ and $\mbox{Im}(\tau_j) > 0$, $d_1$ and
$d_2$ are coefficients of the outgoing waves (also called radiation
coefficients in this paper), $d_{1j}$ and $d_{2j}$ 
are coefficients of the evanescent waves. Since $\mbox{Im}(\alpha) <
0$, the amplitudes of the outgoing waves increase as $|x|$ is
increased. It is well known that resonant modes form bands that depend
on $\beta$ continuously. Each band corresponds to $\omega$ being a
complex-valued function of $\beta$. In the rest of this paper, we
denote a resonant mode by $u_\star$ and its complex frequency by $\omega_\star$.

If the scattering matrix $S$ is invertible,  Eq.~(\ref{Smatrix})
can be written as
\begin{equation}
  \label{Sinv}
  S^{-1} (\omega, \beta) 
  \begin{bmatrix}
    b_1^- \cr b_2^-
  \end{bmatrix}
  =
  \begin{bmatrix}
    b_1^+ \cr b_2^+
  \end{bmatrix}.
\end{equation}
Since the definition of $S$ has been extended to complex
$\omega$, the above is also valid for a resonant mode with a complex
frequency $\omega_\star$. Comparing Eqs.~(\ref{rmode1}) and (\ref{rmode2})
with Eqs.~(\ref{uleft}) and (\ref{uright}), we obtain
\begin{equation}
  \label{Sinvrm0}
  S^{-1} (\omega_\star, \beta) 
  \begin{bmatrix}
    d_1 \cr d_2 
  \end{bmatrix}
  =
  \begin{bmatrix}
    0 \cr 0 
  \end{bmatrix}. 
\end{equation}
Therefore, $S^{-1}$ is singular at $\omega_\star$. In other words,
$\omega_\star$ is a pole of $S$. Using Eq.~(\ref{unitarity}), the
above can be written as
  \begin{equation}
  \label{Sinvrm}
  S^{\sf T} ( \overline{\omega}_\star, \beta) 
  \begin{bmatrix}
    \overline{d}_1 \cr  \overline{d}_2 
  \end{bmatrix}
  =
  \begin{bmatrix}
    0 \cr 0 
  \end{bmatrix}. 
\end{equation}

Due to the reciprocity, corresponding to a resonant mode  $u_\star$ with a real
Bloch wavenumber $\beta \ne 0$ and complex frequency $\omega_\star$, there is
always another resonant mode $u'_\star$ with Bloch wavenumber $-\beta$ and the
same complex frequency $\omega_\star$. Let $d'_1$ and $d'_2$
be the radiation coefficients of $u'_\star$, then
Eq.~(\ref{Sinvrm0}) implies
\[
  S^{-1} (\omega_\star, -\beta) 
  \begin{bmatrix}
    d'_1 \cr d'_2 
  \end{bmatrix}
  =
  \begin{bmatrix}
    0 \cr 0 
  \end{bmatrix}. 
\]
Taking the complex conjugate of above and using Eqs.~(\ref{unitarity})
and (\ref{recip}), we obtain
\begin{equation}
  \label{Satccom}
  S (\overline{\omega}_\star,  \beta) 
  \begin{bmatrix}
    \overline{d}'_1 \cr  \overline{d}'_2 
  \end{bmatrix}
  =
  \begin{bmatrix}
    0 \cr 0 
  \end{bmatrix}. 
\end{equation}
The above means that $\overline{\omega}_\star$ is a zero of the scattering
matrix, i.e.,  for the given $\beta$, $S$ is singular at
$\overline{\omega}_\star$. 
Notice that since $\varepsilon$ is real, $\overline{u}'_\star$  (the
complex conjugate of $u'_\star$) is also a solution of
Eq.~(\ref{helm}). In fact, $\overline{u}'_\star$ is the time reversal of
$u'_\star$. It has a Bloch wavenumber $\beta$,  a complex frequency
$\overline{\omega}_\star$, incoming waves with coefficients
$\overline{d}'_1$
and $\overline{d}'_2$, and no outgoing waves. Equation~(\ref{Satccom})
can be directly obtained by applying $S$ to $\overline{u}'_\star$.

\section{Approximate formulas}

\label{sec:Approximation}

In this section, we derive approximate formulas for a general $2 \times 2$
scattering matrix  and related transmission/reflection spectra,
assuming there is a nondegenerate high quality-factor resonant 
mode with a complex frequency $\omega_\star=\omega_0 - i \gamma$.
The quality factor ($Q$ factor) is given by $Q = \omega_0/(2\gamma)$
and is assumed to be large. The general scattering
matrix $S$ depends on the frequency $\omega$ and satisfies
Eqs.~(\ref{unitarity}), (\ref{Sinvrm}) and (\ref{Satccom}). In
addition, we assume $\omega_\star$ is well separated from other resonances, such that in the
complex $\omega$ plane, there exists a connected domain $\Omega$
containing  $\omega_\star$, $\overline{\omega}_\star$ and $\omega_0$, and $\omega_\star$ is the only pole of $S$ in
$\Omega$. The approximate formulas are valid for $\omega$ near
$\omega_0$.

Since the resonant mode is nondegenerate, $\omega_\star$ is a simple pole
and $\overline{\omega}_\star $ is a simple zero of $S$. Therefore,
\begin{equation}
\label{detS}
\mbox{det}(S) = f(\omega) \frac{\omega -
  \overline{\omega}_\star}{\omega - \omega_\star}, 
\end{equation}
where $f$ is an analytic function of $\omega$ on $\Omega$
and $f(\omega_\star) \neq 0$. Using Eq.~(\ref{unitarity}), it is easy
to show that 
\begin{equation}
    \label{cond_F} \overline{f}(\overline{\omega} ) f(\omega) = 1, 
  \end{equation}
  where $\overline{f}(\overline{\omega})$ is the complex conjugate of
  $f(\overline{\omega})$. Clearly, if $\omega$ is real,
  then $|f(\omega)| = 1$. The function $f$ maps 
$\Omega$ to $f(\Omega) = \{ z = f(\omega) \ | \ \omega \in \Omega
\}$. If in the complex plane, the exterior of $f(\Omega)$ contains a
ray that goes from the origin to infinity, then it can be used as the branch cut
to define a complex square root function, so that $g(\omega)=\sqrt{f(\omega)}$ is
analytic on $\Omega$. Assuming this is the case, we now rewrite the
scattering matrix as  
\begin{equation}
  \label{scaled}
  S(\omega)  =
  \begin{bmatrix}
    r  & \tilde{t} \cr t  & \tilde{r} 
  \end{bmatrix}
  = \frac{g(\omega)}{ \omega - \omega_\star}
  \begin{bmatrix}
    R  & \tilde{T}  \cr T & \tilde{R}
  \end{bmatrix}
\end{equation}
where $R$, $T$, $\tilde{R}$ and $\tilde{T}$ are all analytic functions
of $\omega$ on $\Omega$. 
Using Eqs.~(\ref{unitarity}) and (\ref{detS}), we can show that 
\begin{eqnarray}
  \label{RTprime}
  && \tilde{R}(\omega) = \overline{R} (\overline{\omega}), 
     \qquad 
\tilde{T}(\omega) = - \overline{T} (\overline{\omega} ),  \\
  \label{sum_RT}
 &&  R(\omega) \tilde{R} (\omega)  -  T(\omega) \tilde{T}(\omega)  = (\omega - \omega_\star) (\omega - \overline{\omega}_\star).
\end{eqnarray}

At $\omega_0$, the scattering matrix is 
\begin{equation}
  \label{S0scale}
  S_0 =  S(\omega_0) =
  \begin{bmatrix}
    r_0 & \tilde{t}_0 \cr t_0 & \tilde{r}_0
  \end{bmatrix}
  = \frac{ g_0}{ i \gamma}
  \begin{bmatrix}
    R_0 & \tilde{T}_0 \cr T_0 & \tilde{R}_0
  \end{bmatrix}
\end{equation}
where $r_0  = r (\omega_0)$, $t_0 = t (\omega_0)$, $g_0 = g(\omega_0)$, etc.
From Eqs.~(\ref{sum_RT}) and (\ref{S0scale}), we obtain
\begin{equation}
  \label{FRT0}
  F_0 = g_0^2 = - \det S_0, \quad R_0 =  \frac{i \gamma r_0}{g_0}, \quad 
  T_0 =  \frac{i \gamma t_0}{g_0}.  
\end{equation}
We assume $S_0$ is given  and try to approximate $S$ for $\omega$ near $\omega_0$. For that purpose, we expand $R$
and $T$ in Taylor series at 
$\omega_0$:
\begin{eqnarray}
\label{R_expand} R(\omega)  &=& R_0 + R_1 (\omega - \omega_0) +
          O\left((\omega - \omega_0)^2\right), \\
  \label{T_expand} T(\omega) &=& T_0 + T_1 (\omega - \omega_0 ) +
                                 O\left((\omega - \omega_0)^2\right), 
\end{eqnarray}
where $R_1$ and $T_1$ are the derivatives of $R$ and $T$ (with respect to $\omega$)
evaluated at $\omega_0$. Since $\tilde{R}$ and $\tilde{T}$ satisfy Eq.~(\ref{RTprime}), we have
\begin{eqnarray}
\label{Rpexpand} && \tilde{R}(\omega)  = \overline{R}_0 + \overline{R}_1
                                 (\omega - \omega_0) + 
          O\left((\omega - \omega_0)^2\right), \\
  \label{Tpexpand} && \tilde{T}(\omega) = - \overline{T}_0  -  \overline{T}_1
                                  (\omega - \omega_0 ) + 
                                 O\left((\omega - \omega_0)^2\right). 
\end{eqnarray}
We approximate the scattering matrix by 
\begin{equation*}
  S \approx \frac{g(\omega) }{ \omega - \omega_\star}
  \left\{
    \begin{bmatrix}
      R_0 & -\overline{T}_0 \cr T_0 & \overline{R}_0 
    \end{bmatrix}
    + (\omega-\omega_0)
    \begin{bmatrix}
      R_1 & -\overline{T}_1 \cr T_1 & \overline{R}_1 
    \end{bmatrix}
    \right\}.
\end{equation*}

To find $R_1$ and $T_1$, we use Eq.~(\ref{Sinvrm}) assuming 
${\bf d} = [ d_1, d_2 ]^{\sf T}$ is a given unit
vector. Equation~(\ref{Sinvrm}) can be reduced to
\[
\begin{bmatrix}
  R(\overline{\omega}_\star) & T(\overline{\omega}_\star) \cr 
  - \overline{T}(\omega_\star) & \overline{R}(\omega_\star) 
\end{bmatrix}
\begin{bmatrix}
  \overline{d}_1 \cr \overline{d}_2 
\end{bmatrix}
=
\begin{bmatrix}
0 \cr 0 
\end{bmatrix}. 
\]
Writing down the above using the expansions of $R$ and $T$, we obtain
\begin{eqnarray}
  \label{R1sol}
  && R_1  \approx \frac{ 1}{g_0} \left[ ( |d_2|^2 - |d_1|^2) r_0 - 2 
     d_1 \overline{d}_2 t_0 \right] \\
  \label{T1sol}
  && T_1  \approx \frac{ 1}{g_0} \left[       - 2  \overline{d}_1  d_2
     r_0 + ( |d_1|^2 - |d_2|^2) t_0  \right].
\end{eqnarray}
The above can be written as
\[
  \begin{bmatrix}
    R_1 \cr T_1 
  \end{bmatrix}
  \approx \frac{1}{g_0} H
  \begin{bmatrix}
    r_0 \cr t_0 
  \end{bmatrix}
\]
where $H = I - 2 {\bf d} {\bf d}^*$ is a
Hermitian unitary matrix satisfying $H = H^* = H^{-1}$.
Let $\rho(\omega) = g_0 / g(\omega)$, then the 
final result is 
\begin{eqnarray}
 \nonumber  \rho(\omega)  S(\omega) & \approx & S_0 -
  2 \frac{ \omega - \omega_0}{\omega 
    - \omega_\star} {\bf d} {\bf p}^{\sf T} \\
  \label{rhoS}
  &=& \left( I - 2 \frac{ \omega - \omega_0}{\omega 
    - \omega_\star} {\bf d} {\bf d}^* \right) S_0, 
\end{eqnarray}
where ${\bf p}  = S_0^{\sf T} \overline{\bf d}$ and
$I$ is the identity matrix. 

Equation~(\ref{rhoS}) approximates $\rho(\omega) S(\omega)$
using the scattering matrix at $\omega_0$, the complex frequency
$\omega_\star$ and the radiation coefficients ${\bf d}$ of the 
resonant mode. However, it is 
not an approximation to $S$, since $\rho$ is an unknown function
related to $f$. Fortunately, for any real $\omega$, $|\rho(\omega)|=1$,
thus, the reflection and transmission spectra can 
be approximated precisely. The first column of Eq.~(\ref{rhoS}) gives
\begin{eqnarray}
  \label{rabs}
&&  |r(\omega)| \approx \left| r_0 - 2 \frac{ \omega - \omega_0}{\omega 
  - \omega_\star} \left( |d_1|^2 r_0 + d_1 \overline{d}_2 t_0 \right) 
   \right|, \\
  \label{tabs}
&&  |t(\omega)| \approx \left| t_0 - 2 \frac{ \omega - \omega_0}{\omega 
  - \omega_\star} \left( \overline{d}_1 d_2 r_0 + |d_2|^2  t_0 \right) 
  \right|.
\end{eqnarray}
Moreover, Eq.~(\ref{rhoS}) allows us to find approximately the zeros of the
transmission and reflection coefficients. 
Let $\omega^\circ_{r}$ and $\omega^\circ_{t}$ be the zeros of
$r(\omega)$ and $t(\omega)$, respectively. For simplicity, we call
$\omega_r^\circ$ a reflection zero and $\omega_t^\circ$ a transmission
zero. 
From the leading terms in (\ref{R_expand}) and
(\ref{T_expand}), and assuming $R_1$ and $T_1$ are nonzero, we get 
\[
  \omega^\circ_r \approx \omega_0 - \frac{R_0}{R_1}, \quad
  \omega^\circ_t \approx \omega_0 - \frac{T_0}{T_1}.
\]
Using  $R_0$, $T_0$, $R_1$ and $T_1$  given in Eqs.~(\ref{FRT0}),
(\ref{R1sol}) and (\ref{T1sol}), we obtain
\begin{eqnarray}
\label{rzero}
&&  \omega^\circ_{r} \approx \omega_0 + \frac{ i \gamma r_0}
   { ( |d_1|^2 - |d_2|^2) r_0 + 2 d_1 \overline{d}_2 t_0}, \\
\label{tzero}
  &&  \omega^\circ_{t} \approx \omega_0 + \frac{ i \gamma t_0}
   {   2 \overline{d}_1 d_2 r_0 + ( |d_2|^2 - |d_1|^2) t_0 }. 
\end{eqnarray}
Apparently, $\omega^\circ_r$ and $\omega_t^\circ$ are complex in
general.

In Sec.~II, we mentioned that when the periodic structure has a proper
symmetry, the reflection and/or transmission coefficients for the left and
right incident waves are identical, and in that case, $\omega^\circ_r$
and/or $\omega^\circ_r$ are real~\cite{popov86,gipp05}. 
For the case of equal transmission coefficients, i.e., $t = \tilde{t}$ for
all $\omega$, the scattering matrix $S$ is symmetric, thus $T(\omega)
= - \overline{T}(\overline{\omega} )$. This implies that if
$\omega$ is real, then $T(\omega)$ is pure imaginary, and
consequently, $T_0$ and $T_1$ are pure imaginary, and 
$\omega^\circ_t$ is real. Considering the leading terms in the expansions
(\ref{R_expand})-(\ref{Tpexpand}), we have 
\[
    \frac{t_0}{g_0}  = \overline{
      \left(\frac{t_0}{g_0} \right) },  
    \quad 
  \frac{ \tilde{r}_0}{g_0}  = - \overline{\left(
    \frac{r_0}{g_0}  \right)}. 
\]
Using $T_1$ given in Eq.~(\ref{T1sol}) and the condition $T_1 = -
\overline{T}_1$, we obtain 
\[
  |d_1|^2 t_0 + d_1 \overline{d}_2 \tilde{r}_0 
  = \overline{d}_1 d_2 r_0 + |d_2|^2 t_0.
\]
The above implies that ${\bf d}{\bf p}^{\sf T}$ is a symmetric matrix, thus
  the right hand side of Eq.~(\ref{rhoS}) is symmetric. In addition,
  Eq.~(\ref{tzero}) can be written as
  \begin{equation}
    \label{realtzero}
    \omega^\circ_t \approx \omega_0 +  \frac{ \gamma t_0/ g_0} {
      2 \mbox{Im} ( \overline{d}_1 d_2 r_0/g_0 ) }.
  \end{equation}
The above gives an approximate real zero for the
transmission coefficient. Notice that the above formula requires a
nonzero $r_0$.

For the case of equal reflection coefficients, i.e. $r = \tilde{r}$
for all $\omega$, we have $R(\omega) =
\overline{R}(\overline{\omega})$. Therefore,   $R(\omega)$ is real for real
$\omega$, and $R_0$ and $R_1$ are also real. The leading terms in the
expansions
(\ref{R_expand})-(\ref{Tpexpand}) give rise to
\[
    \frac{r_0}{g_0}  = - \overline{
      \left(\frac{r_0}{g_0} \right) },  
    \quad 
  \frac{ \tilde{t}_0}{g_0}  =  \overline{\left(
    \frac{t_0}{g_0}  \right)}. 
\]
The condition $R_1 = \overline{R}_1$ leads to
\[
  |d_1|^2 r_0 + d_1 \overline{d}_2 t_0 
  = \overline{d}_1 d_2 \tilde{t}_0 + |d_2|^2 r_0.
\]
The above implies that the $(1,1)$ and $(2,2)$ entries of matrix ${\bf
  d}{\bf p}^{\sf T}$, thus  the right hand side of Eq.~(\ref{rhoS}), are
the same. Moreover, Eq.~(\ref{rzero}) can be written as
  \begin{equation}
    \label{realrzero}
    \omega_r^\circ \approx \omega_0 + \frac{ i \gamma r_0/
      g_0 } { 2 \mbox{Re} ( d_1 \overline{d}_2 t_0/g_0 ) }
  \end{equation}
  and $\omega_r^\circ $ is real. 

For the case with $t=\tilde{t}$ and $r = \tilde{r}$,
Eq.~(\ref{Sinvrm}) becomes 
\[
\begin{bmatrix}
  R(\overline{\omega}_\star) & T(\overline{\omega}_\star) \cr
 T(\overline{\omega}_\star) &     R(\overline{\omega}_\star) 
\end{bmatrix}
\begin{bmatrix}
  \overline{d}_1 \cr \overline{d}_2 
\end{bmatrix}
=
\begin{bmatrix}
0 \cr 0 
\end{bmatrix}. 
\]
Since the resonant mode with 
complex frequency $\omega_*$ is nondegenerate,
$R(\overline{\omega}_\star)$ and $T(\overline{\omega}_\star)$ 
cannot both be zero. It is also impossible for one of them to be zero,
because otherwise, ${\bf d}$ would be a zero vector.  Therefore, both
$R(\overline{\omega}_\star)$ and 
$T(\overline{\omega}_\star)$ are nonzero. In that case, $d_1^2 =
d_2^2$, and  we can scale ${\bf d}$, such that
\[
  d_1 = \pm d_2 = 1/\sqrt{2},
\]
where the plus or minus sign depends on the symmetry of the resonant
mode. With the given ${\bf d}$, the formulas for $R_1$ and $T_1$ are
simplified to 
\[
  R_1 \approx \mp \frac{ t_0}{g_0}, \quad
  T_1 \approx \mp \frac{r_0}{g_0}. 
\]
Equations (\ref{rabs}) and (\ref{tabs}) are reduced to
\begin{eqnarray}
\label{rrabs} &&  |r(\omega)| \approx  \left|  \frac{  i \gamma r_0 \mp (\omega - \omega_0) 
   t_0  }{ \omega - \omega_\star } \right|, \\
\label{ttabs} &&  |t(\omega)| \approx  \left|  \frac{  i \gamma t_0 \mp (\omega - \omega_0) 
   r_0  }{ \omega - \omega_\star } \right|.
\end{eqnarray}
Assuming both $r_0$ and $t_0$ are nonzero, we can simplify the
expressions for the zeros of the reflection and transmission
coefficients as
\begin{equation}
  \label{realzeros2}
  \omega_r^\circ \approx \omega_0 \pm \frac{ i \gamma r_0}{t_0}, \quad 
  \omega_t^\circ \approx \omega_0 \pm \frac{ i \gamma t_0}{r_0}.
\end{equation}
Since $t_0/g_0$ is real and $r_0/g_0$ is pure imaginary,
$t_0/r_0$ and $r_0/t_0$ are pure imaginary. Therefore, both $\omega_r^\circ$ and
$\omega_t^\circ$ are real.

\section{Coupled mode theory}

In a seminal work~\cite{fan03}, Fan {\it et al.} developed a TCMT for
a resonator connected with $m$
ports. Assuming the resonator has a single resonant mode with a complex
frequency $\omega_\star = \omega_0 - i \gamma$ and there is no material loss in the structure, the TCMT states that
\begin{eqnarray}
  \label{cmta}
 \frac{d a}{d {\sf t}} &=& - i \, \omega_\star a + {\bm p}^{\sf T} {\bm b}^+  \\
  \label{cmtb}
  {\bm b}^-   &=&  C {\bm b}^+ + a {\bm d} \\
  \label{dnorm}
   || {\bm d} ||^2 &=& 2 \gamma \\
  \label{cmtcc}
 C^* &=&  C^{-1}   \\
\label{cmtdd}
   {\bm p}  &=& -    C^{\sf T} \overline{\bm d} \\
 C &=& C^{\sf T} \\
   {\bm p} &=& \label{cmtd}  {\bm d} 
\end{eqnarray}
where  $a=a({\sf t})$ is time-dependent amplitude of the resonant mode
scaled such that $|a|^2$ is the energy of the resonant mode in the
resonator, ${\bm b}^+$ and ${\bm b}^-$ are column vectors of $b_j^+$
and $b_j^-$ (for $j=1$, 2, ..., $m$), respectively,
$b_j^+ = b_j^+({\sf t})$ is the time-dependent amplitude of the incoming
wave in the $j$th port scaled such that $|b_j^+|^2$ is the power of the incoming wave,
$b_j^-$ is similarly defined for the outgoing wave, ${\bm p}$ is
a column vector of coupling coefficients connecting incoming waves with
the resonant mode, ${\bm d}$ is a column vector for the radiation
coefficients of the resonant mode and it couples the resonant mode  to
the outgoing waves, $C$ is the scattering matrix
for the direct non-resonant passway.
For time harmonic waves, the TCMT gives the
following scattering matrix:
\begin{equation}
  \label{tcmtSmatrix}
  S(\omega) = C - \frac{ {\bm d} {\bm p}^{\sf T} }{ i 
    (\omega - \omega_\star)}.
\end{equation}

The above TCMT for a single-mode resonator is constructed by
considering energy conservation, reciprocity and time-reversal
symmetry. It is assumed that the original and reciprocal waves exist
in the same resonator/ports structure, and consequently, $C$ and $S$
are required to be symmetric. Since energy must be conserved, the
matrix $C$ should be unitary. Additional conditions on ${\bm d}$ and
${\bm p}$, including Eq.~(\ref{dnorm}), are obtained when
Eqs.~(\ref{cmta}) and (\ref{cmtb}) are applied to the resonant mode
and its time reversal. These conditions and the symmetry of $S$ give
rise to Eq.~(\ref{cmtd}).  Equation~(\ref{cmtdd}) is obtained when the
scattering matrix is applied to the time-reversed resonant mode.
The TCMT can be extended to more complicated resonant systems. 
A TCMT for multimode resonators was developed by Suh {\it et
  al.}~\cite{fan04}. Recently, Zhao {\it et 
  al.}~\cite{zhao19} developed a new TCMT by considering both the
original physical system and the time-reversal conjugate system. The
new TCMT establishes the constraints of energy conservation, reciprocity 
and time-reversal symmetry separately, and it is applicable to a wider
range of resonant systems. For reciprocal systems, the scattering
matrices $C$ and $S$ are also symmetric in the recent
works~\cite{wang18,zhao19}.

The TCMT is applicable to diffraction problems of periodic 
structures with normal incident plane waves where the ports are 
the propagating diffraction orders. However, it is not applicable to
diffraction problems with oblique incident waves, since in that case
the scattering matrix is not symmetric. As we mentioned in Sec.~II,
when there is a nonzero 
wavenumber $\beta$ for the periodic direction, the reciprocal wave
has a different set of diffraction orders, and the scattering matrix
satisfies Eq.~(\ref{recip}) and is non-symmetric in
general. Furthermore,  to apply
the TCMT to a specific problem, it is necessary to calculate the complex
frequency $\omega_\star$ and radiation coefficients ${\bm d}$, and
estimate the scattering matrix $C$. It appears that $C$
cannot be calculated rigorously, because the resonant and non-resonant
wave field components cannot be separated easily. For the case of a photonic
crystal slab, the matrix $C$ may be approximated by the
scattering matrix of a uniform slab, but the refractive index 
of the uniform slab can only be obtained by data fitting~\cite{fan03,fan02}.

In the following, we present a revised TCMT where the scattering
matrix is non-symmetric in general and $C$ is replaced by the
scattering matrix at $\omega_0$.
We start with the same Eqs.~(\ref{cmta}) and (\ref{cmtb}) for a resonant mode
with amplitude $a$ and complex frequency $\omega_\star$, incoming/outgoing
waves with amplitudes $b_j^\pm$ in the $j$th radiation channel, and a scattering matrix $C$ for the direct non-resonant
passway. The scattering matrix given in Eq.~(\ref{tcmtSmatrix})
remains valid. Since the reciprocal waves propagate in  a different set of
radiation channels, we have
\begin{eqnarray}
  \label{rcmta}
  \frac{d a'}{d {\sf t}} &=& - i \, \omega_\star a' + {\bm p}^{\prime\sf
                             T} {\bm b}^{\prime +}  \\
  \label{rcmtb}
  {\bm b}^{\prime -}   &=&  C' {\bm b}^{\prime +} + a'  {\bm d}'
   \\
  \label{rcmtSmatrix}
  S'(\omega) &=& C'- \frac{ {\bm d}' {\bm p}^{\prime \sf
                       T} }{ i     (\omega - \omega_\star)}
\end{eqnarray}
where $a'$ is the amplitude of the reciprocal mode, $b_j^{\prime +}$
is the amplitude of incoming wave in the $j$th reciprocal radiation
channel, $C'$ is the scattering matrix for direct passway in the
reciprocal system, $S'$ is the frequency-dependent scattering
matrix of the reciprocal system, etc. Notice that 
  Eqs.~(\ref{rcmta})-(\ref{rcmtSmatrix}) are different
  from those for the time-reversal conjugate system~\cite{zhao19}.

In view of Eq.~(\ref{recip}), the reciprocity principle requires that
$C' = C^{\sf T}$ and $S' = S^{\sf T}$, and
thus
\begin{equation}
  \label{dksym}
  {\bm d}' {\bm p}^{\prime \sf T}
    = {\bm p} {\bm d}^{\sf T}.  
\end{equation}
In addition, the conservation of energy implies that $C$ must be
a unitary matrix. Applying the theory to the resonant mode and the
reciprocal mode as in Ref.~\cite{fan03}, we obtain
\begin{equation}
  \label{normd}
  || {\bm d} ||^2 = || {\bm d}' ||^2 = 2 \gamma.
\end{equation}
Importantly, the time-reversed resonant mode propagates in the
reciprocal radiation channels, and satisfies
Eqs.~(\ref{rcmta})-(\ref{rcmtSmatrix}), and the time-reversed
reciprocal mode satisfies Eqs.~(\ref{cmta}), (\ref{cmtb}) and 
(\ref{tcmtSmatrix}). Applying the theory to the time-reversed modes as in
Ref.~\cite{fan03}, we obtain
\begin{eqnarray}
  \label{normk}
&&   {\bm p}^{\sf T} \overline{ {\bm d}^{\prime} }
   =   {\bm p}^{\prime \sf T} \overline{\bm d} = 2 \gamma, \\
  \label{dCd}
  && C \overline{ {\bm d}' } = - {\bm d}, \quad C^{\sf T}
     \overline{\bm d} = - {\bm d}'. 
\end{eqnarray}
Solving Eqs.~(\ref{dksym}), (\ref{normd}) and (\ref{normk}), we obtain
\begin{equation}
  \label{keqd}
  {\bm p}  = {\bm d}', \quad
  {\bm p}'   = {\bm d}.   
\end{equation}
Therefore, ${\bm p}$ is the vector of radiation coefficients of
the reciprocal mode,  Eq.~(\ref{cmtdd}) is still valid,  and 
\begin{equation}
  \label{Smat2}
  S(\omega) = \left[ I + \frac{ {\bm d} {\bm d}^* }{ i
      (\omega - \omega_\star) } \right] C. 
\end{equation}

In summary, if the original and reciprocal waves propagate in
different radiation channels, then TCMT should use
Eqs.~(\ref{cmta})-(\ref{cmtdd}). The scattering matrix is given
in Eq.~(\ref{tcmtSmatrix}) or (\ref{Smat2}).

At $\omega=\omega_0$, Eq.~(\ref{tcmtSmatrix}) becomes
\begin{equation}
  \label{cmtS0}
S(\omega_0)=   S_0 = C + \frac{1}{\gamma} {\bm d}{\bm p}^{\sf T}.
\end{equation}
Therefore,
\begin{equation}
  \label{cmtS3}
  S(\omega) = S_0 - \frac{ \omega - \omega_0}{ \gamma 
    (\omega - \omega_\star)} {\bm d}{\bm p}^{\sf T}. 
\end{equation}
Multiplying ${\bm d}^*$  to both sides of
Eq.~(\ref{cmtS0}), we obtain   
\begin{equation}
  {\bm p}^{\sf T} = {\bm d}^*  S_0.  
\end{equation} 
Using the unit vectors
\begin{equation}
  \label{unitvec}
  {\bf d} = \frac{ \bm d}{ || {\bm d} ||}, \quad
  {\bf p}  = \frac{ {\bm p}} { || {\bm p} ||}
  = S_0^{\sf T} \overline{\bf d}, 
\end{equation}
we can rewrite the scattering matrix as 
\begin{eqnarray}
\nonumber 
  S(\omega) &=& S_0 -
  2 \frac{ \omega - \omega_0}{ \omega -
    \omega_\star} {\bf d}{\bf p}^{\sf T} \\
  \label{cmtS4}  &=& \left( I - 2 \frac{ \omega - \omega_0}{ \omega -
    \omega_\star} {\bf d}  {\bf d}^*  \right) S_0.
\end{eqnarray}
It can be easily verified that 
\begin{equation}
S (\omega) S^*(\overline{\omega}) = I. 
\end{equation}
Therefore, if $\omega$ is real, $S$ is unitary. Moreover,
\[
  S^{-1}(\omega_\star) {\bf d} =
  S^*(\overline{\omega}_\star)   {\bf d} = S_0^* ( I - {\bf d} {\bf d}^* ) {\bf d}
  = {\bf 0}. 
\]
Therefore, $\omega_\star$ is a zero of $S^{-1}$ and a pole of ${\bf
  S}$. Similarly, 
\[
  S^{-{\sf T}}(\omega_\star) {\bf p} =
  \overline{S}(\overline{\omega}_\star) {\bf p} = 
  ( I - \overline{\bf d} {\bf d}^{\sf T} )  \overline{S}_0  {\bf p}
= {\bf 0}. 
\]

Notice that Eq.~(\ref{cmtS4}) is similar but not identical to  Eq.~(\ref{rhoS}) in
Sec.~III. The latter is derived from the exact scattering
matrix, but it is only valid for the $2 \times 2$ case and it contains an
unknown analytic function $\rho$ satisfying $\rho(\omega_0)=1$.
For $\omega$ near $\omega_0$,  if  we approximate $\rho(\omega)$ by 1,
then Eq.~(\ref{rhoS}) is reduced to Eq.~(\ref{cmtS4}). It should be
emphasized that Eq.~(\ref{cmtS4}) is only a model. Although TCMT follows  the most important physical
principles, it ignores the coupling caused by the evanescent waves,
ignores the frequency dependence of the incoming and outgoing waves and the coupling coefficients, ignores the difference between the
actual field in the resonator and the resonant mode, etc.  On the
other hand, Eqs.~(\ref{rhoS}) and (\ref{cmtS4}) do give the same approximate zeros
of the reflection and transmission coefficients, and since 
$|\rho(\omega)|=1$ for  real $\omega$, they also give the same
approximate transmission and reflection spectra.

\section{Numerical Examples}
\label{sec:examples}
\begin{figure}[http]
  \centering 
   \includegraphics[scale=0.7]{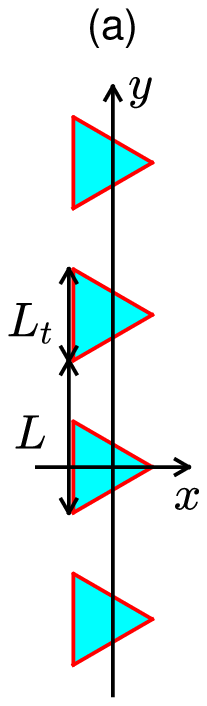}
    \includegraphics[scale=0.7]{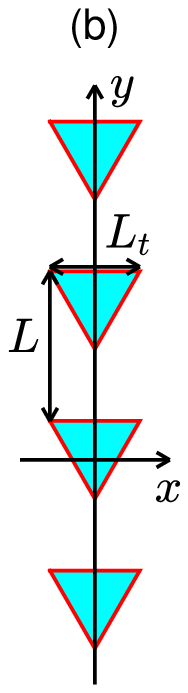}
   \includegraphics[scale=0.7]{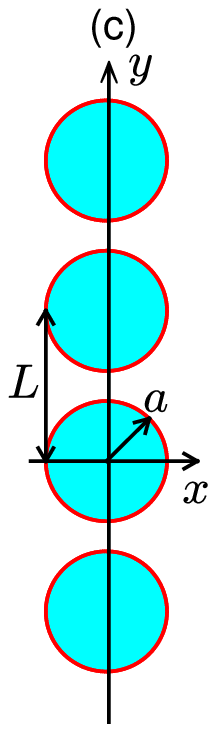}
 \caption{Three periodic arrays of cylinders with period $L$ in the
   $y$ direction. The cylinders have three different shapes: (a):
   equilateral triangles with a reflection symmetry in $y$, (b):
   equilateral triangles with a reflection symmetry in $x$, (c):
   circular cylinders.} 
 \label{fig_stru}
\end{figure}

In this section,  we present numerical examples to validate the
approximate formulas derived in Sec.~\ref{sec:Approximation}. The
numerical results are obtained for three periodic arrays of dielectric
cylinders shown in Fig.~\ref{fig_stru}.
The arrays are periodic in $y$ with period $L$ and the cylinders are surrounded by air.
The dielectric  constants  of the cylinders and surrounding air are
$\varepsilon_1=10$ and $\varepsilon_0=1$, respectively. The cross
sections of the cylinders shown in Fig.~\ref{fig_stru}(a) and (b) 
are equilateral triangles with side length $L_t$. The
radius of the circular cylinders shown in Fig.~\ref{fig_stru}(c) is 
$a$. The arrays with triangular cylinders have a reflection symmetry in $y$ or $x$.
The array with circular cylinders is symmetric in both $x$ and $y$.

\begin{figure}[http]
    \centering 
    \includegraphics[scale=0.79]{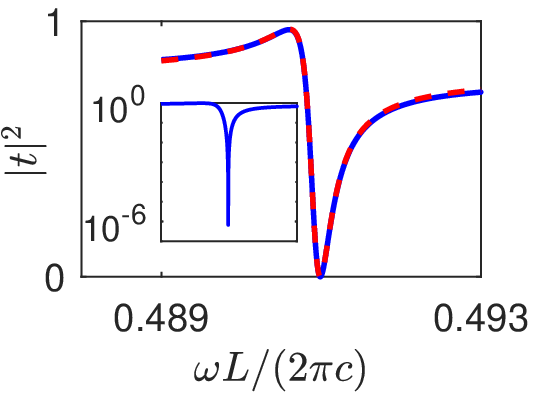}
     \includegraphics[scale=0.79]{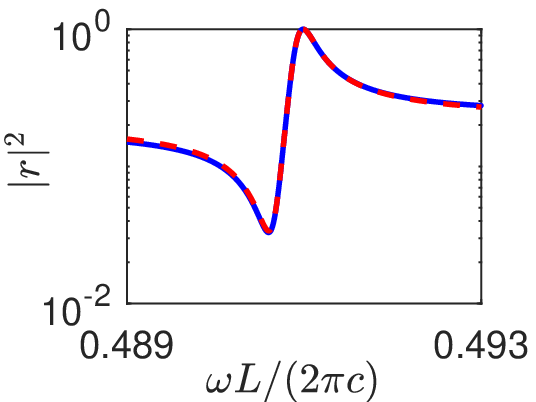}
   \caption{Transmission and reflection spectra near a resonant
     frequency for a periodic array shown in
     Fig.~\ref{fig_stru}(a). The results are obtained for 
     $L_t = 0.7L$ and $\beta = 0.02 (2\pi/L)$. The inset shows the
     transmission spectrum in 
     a logarithmic scale. The numerical and approximate analytic
     results are shown as the solid blue lines and dashed red lines,
     respectively.}
   \label{fig_RES1}
 \end{figure}

Resonant modes in the periodic arrays form bands that depend on the
real Bloch wavenumber $\beta$ continuously.
For $\beta = 0.02\, (2\pi/L)$ and $L_t = 0.7L$, the
periodic array shown in Fig.~\ref{fig_stru}(a) supports a 
resonant mode with normalized complex frequency 
$\omega_\star L/(2\pi c) = 0.49092 - 1.51 \times 10^{-4} i$
and radiation coefficients satisfying
$ d_1/d_2 = 0.8281 - 0.0696 i$.
For  the real frequency $\omega_0 = \mbox{Re}(\omega)$, we solve the Helmholtz 
equation (\ref{helm}) numerically and obtain the scattering matrix $S_0$.
The reflection and transmission coefficients $r_0$ and $t_0$ (for left incident
waves) satisfy $|r_0|^2 = 0.821$ and $|t_0|^2 = 0.179$. 
In Fig.~\ref{fig_RES1},
 we show the transmission and reflection spectra for the same $\beta$
 and for $\omega$ near $\omega_0$. 
 The solid blue lines and dashed red lines
  correspond to results obtained by numerical simulation and the
  approximate formulas (\ref{rabs}) and (\ref{tabs}), respectively. 
  The numerical and analytic results agree very well.  The
  transmission coefficient has a real zero $\omega_t^{\circ} \approx
  0.49099 (2\pi c/L)$. The approximate  formula (\ref{tzero}) or
  (\ref{realtzero}) gives $\omega_t^\circ$ with five correct digits. 
  Since the periodic structure has only a reflection symmetry in $y$,
  the zero of the reflection coefficient is complex, and the
  reflection spectrum has a nonzero dip. 

\begin{figure}[http]
      \centering 
      \includegraphics[scale=0.79]{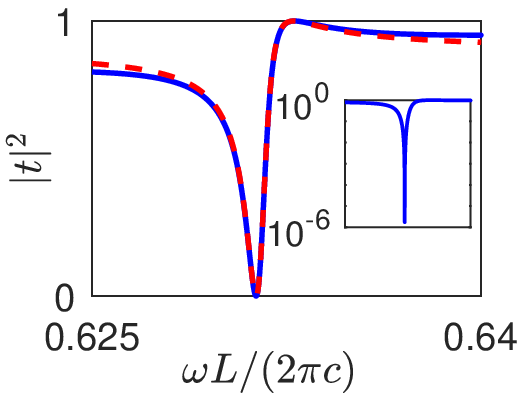}
       \includegraphics[scale=0.79]{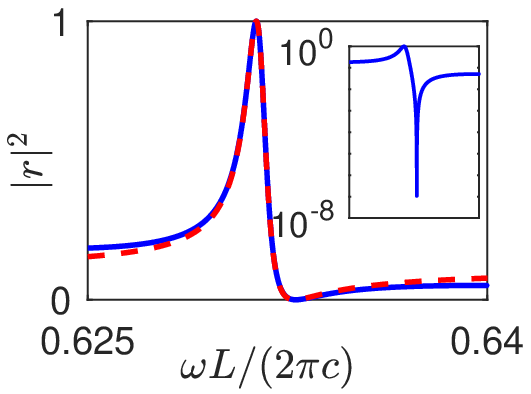}
     \caption{Transmission and reflection spectra near a resonant
       frequency for the periodic array shown in
       Fig.~\ref{fig_stru}(b).  The results are obtained for $L_t =
       0.5L$ and $\beta = 0.02 (2\pi/L)$. The insets 
       show the spectra in a logarithmic scale. Numerical and
       approximate analytic results are shown as  
       the solid blue lines and the dashed red lines, respectively.}
     \label{fig_RES2}
      \end{figure}

  For the periodic array shown in Fig.~\ref{fig_stru}(b) with $L_t =
  0.5L$ and $\beta  = 0.02\, (2\pi/L)$, we found 
a resonant mode with normalized complex frequency
$\omega_\star L/(2\pi c) = 0.63148 - 4.49 \times 10^{-4}  i $. This mode is even in $x$, and thus the radiation coefficients
are $d_1 = d_2 = 1/\sqrt{2}$. At $\omega_0$, we found reflection and
transmission coefficients satisfying
$|r_0|^2 = 0.892$ and $|t_0|^2 = 0.108$. In Fig.~\ref{fig_RES2},
      we show transmission and reflection spectra for frequencies near
      $\omega_0$. The numerical results are shown as the 
      solid blue lines, and compared with the analytic
      approximations shown as the dashed red lines. A very good
      agreement is achieved. The approximate
      results are calculated by the formulas (\ref{rrabs}) and
      (\ref{ttabs}) with ``$\mp$'' replaced by the minus sign. 
      Since the periodic structure is symmetric in $x$,
      the transmission and reflection coefficients have real zeros
      $\omega_t^\circ \approx 0.63133 (2\pi c/L)$ and $\omega_r^\circ =
      0.63281 (2\pi c/L)$, respectively. The approximate formula
      (\ref{realzeros2}) gives $\omega_t^\circ$ with the same five 
      digits and a real reflection zero $\omega_r^\circ
      \approx 0.63277 (2\pi c/L)$ (with four correct digits after
      rounding). 

\begin{figure}[http]
  \centering 
  \includegraphics[scale=0.8]{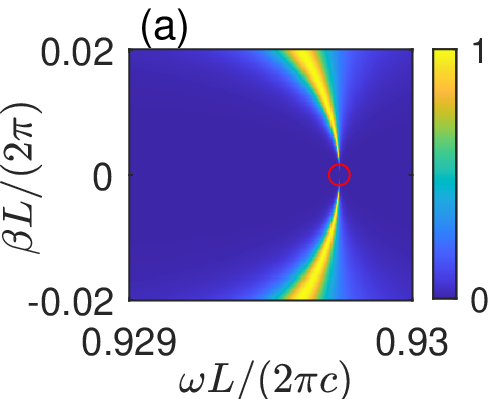}
  \includegraphics[scale=0.8]{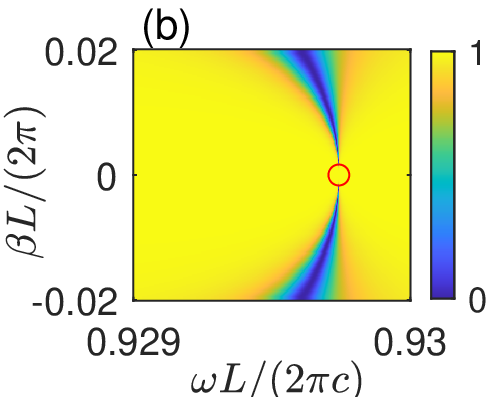}
  \includegraphics[scale=0.8]{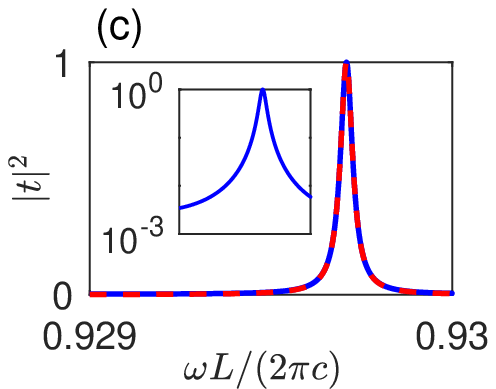}
  \includegraphics[scale=0.8]{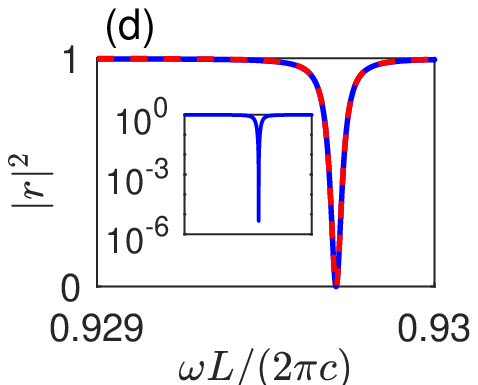}
  \caption{Transmittance and reflectance near a BIC, marked as a small 
    circle in (a) and (b), for a periodic array of 
    circular cylinders (radius $a=0.2694L$) shown in 
    Fig.~\ref{fig_stru}(c). 
    (a) Transmittance as a function of $\omega$ and $\beta$; 
    (b) Reflectance  as a function of $\omega$ and $\beta$; 
    (c) Transmittance for fixed $\beta = 0.01 (2\pi/L)$; 
    (d) Reflectance for fixed $\beta = 0.01 (2\pi/L)$. 
    In (c) and (d), the numerical and approximate analytic results are shown as the 
    solid blue lines and dashed red lines, respectively. 
    The insets show the spectra in a logarithmic scale.}
  \label{fig_ASW2}
\end{figure}

In Sec.~III, we showed that a proper symmetry and a nonzero value of
$r_0$ (or $t_0$) are conditions for the existence of a real transmission
(or reflection) zero. To illustrate this, we consider the periodic array
of circular cylinders shown in Fig.~\ref{fig_stru}(c). It is well-known that a lossless periodic
dielectric array can support a variety of bound states in the
continuum (BICs) which are 
special resonant modes with a real frequency and an infinite $Q$
factor~\cite{bulg14,hu15,hsu16,yuanJPB,kosh19,jin19,amgad21,luo21,sad21}. 
For the cylinder radius $a = 0.2694L$, the periodic array has
a symmetry-protected BIC with wavenumber $\beta_\dagger=0$ and frequency $\omega_\dagger = 0.9297
(2\pi c/L)$. The radius $a$ is 
chosen so that the transmission coefficient (for
$\beta=0$) at the BIC frequency $\omega_\dagger$ is exactly zero.
To understand the transmission and reflection spectra
  for $\beta$ near $\beta_\dagger$, it is necessary to consider $t$
  and $r$ as functions of two variables 
  $\omega$ and $\beta$.
  In Fig.~\ref{fig_ASW2}(a) and (b),
 we show transmittance $|t|^2$ and 
reflectance $|r|^2$ as functions of $\omega$ and $\beta$, 
respectively.   It is known that $t$ and $r$ (as functions of
two variables) are discontinuous at
$(\omega_\dagger, \beta_\dagger)$. For this example, although
$t(\omega_\dagger, \beta_\dagger) = 0$ and
$|r(\omega_\dagger,\beta_\dagger)|=1$, there is a function of $\beta$,
namely $\omega_r^\circ = \omega_r^\circ (\beta)$, such that
$r( \omega_r^\circ, \beta) = 0$ and $|t( \omega_r^\circ, \beta)| = 1$.
Meanwhile, for $\beta$ near $\beta_\dagger=0$, there is a resonant mode with a
  complex frequency $\omega_\star= \omega_\star(\beta)$, so that 
  $\omega_\star (\beta_\dagger) = \omega_\dagger$.
It turns out that $\omega_r^\circ \approx \omega_0 =
\mbox{Re}(\omega_\star)$ for $\beta$ near $\beta_\dagger=0$.
Specifically, for $\beta = 0.01 (2\pi/L)$, the normalized complex
frequency of the resonant mode is
$\omega_\star L/(2\pi c)  = 0.92965 - 2.1 \times 10^{-5} i$.
The reflection  coefficient $r_0 = r (\omega_0, \beta)$ satisfies $|r_0|^2 = 5.9 
\times 10^{-7}$ and it is close to zero. Therefore, the formula for  
$\omega_t^\circ$ in Eq.~(\ref{realzeros2}) breaks down, and there is no real
transmission zero near $\omega_0$. In Fig.~\ref{fig_ASW2}(c) and (d), 
we show the transmission and reflection spectra for $\beta = 0.01
(2\pi/L)$. The transmission spectrum has a Lorentzian line shape with
a $100\%$ peak, and it does not reach zero. The reflection spectrum
has a zero dip at $\omega_r^\circ \approx \omega_0$. The solid blue lines shown in
Fig.~\ref{fig_ASW2} are the numerical results. Analytic results based
on Eqs.~(\ref{rrabs}) and (\ref{ttabs}) are shown as the dashed red
lines, and they agree with the numerical results very well. Since the
resonant mode is even in $x$. The ``$\mp$'' signs in
Eqs.~(\ref{rrabs}) and (\ref{ttabs}) are replaced by the minus sign.

\section{Conclusion}
\label{sec:conclusions}

For structures with a high-$Q$ resonant mode, wave scattering exhibits
interesting resonance phenomena with sharp peaks and/or  
dips in transmission, reflection and other spectra. Analytic studies or
models are useful, because numerical solutions are expensive to
obtain and do not provide much physical insight.
For scattering problems with two radiation channels and
assuming the existence of a nondegenerate high-$Q$ resonant mode 
sufficiently separated from other resonances, we derived approximate
formulas (for the scattering matrix and transmission/reflection
spectra) directly from the exact scattering matrix. Unlike the
existing model of Popov {\it et al.}~\cite{popov86},
we do not need to solve the governing PDE to find 
the zeros of the transmission/reflection coefficients.
In fact,  our approximate formulas predict the transmission and
reflection zeros, whether they are real or complex. 

Constructed from a few basic physical principles, the TCMT of Fan {\it
  et al.}~\cite{fan03} gives a symmetric scattering-matrix model that depends on the scattering 
matrix $C$ for the direct non-resonant passway. The model is
simple and elegant, but $C$ cannot be calculated rigorously. We revised the
TCMT to scattering problems with (in general) non-symmetric 
scattering matrices and replaced $C$ by the 
scattering matrix $S_0$ at the (real) resonant
frequency. The revised TCMT and the theory based on direct derivation
lead to slightly different approximations to the 
scattering matrix, but they give exactly the same transmission and
reflection spectra. The directly derived results are rigorous, can
be further improved if more terms in the Taylor series (\ref{R_expand}) and
(\ref{T_expand}) are included, but they are restricted to $2 \times 2$
scattering matrices. It is worthwhile to further extend the theories developed
in this paper, for example, to problems with a few interacting and possibly
degenerate resonant modes.

\section*{Acknowledgement}
 The authors acknowledge support from the Natural Science Foundation
 of Chongqing, China (Grant No. cstc2019jcyj-msxmX0717),   and the
 Research Grants Council of Hong Kong Special Administrative Region,
 China (Grant No. CityU 11305518).

\end{document}